# Sequential & Parallel Algorithms
# For the Addition of Big-Integer Numbers

Youssef Bassil , Aziz Barbar

American University of Science & Technology, Lebanon
{ybassil , abarbar}@aust.edu.lb

**Abstract.** Today's PCs can directly manipulate numbers not longer than 64 bits because the size of the CPU registers and the data-path are limited. Consequently, arithmetic operations such as addition, can only be performed on numbers of that length. To solve the problem of computation on big-integer numbers, different algorithms were developed. However, these algorithms are considerably slow because they operate on individual bits; and are only designed to run over single-processor computers. In this paper, two algorithms for handling arithmetic addition on big-integer numbers are presented. The first algorithm is sequential while the second is parallel. Both algorithms, unlike existing ones, perform addition on blocks or tokens of 60 bits (18 digits), and thus boosting the execution time by a factor of 60.

## 1 Introduction

Modern PCs are very good at handling and doing math on numbers whose length does not exceed 32 bits or 64 bits. [1] However, when numbers become larger than that, computer arithmetic, such as addition, subtraction, multiplication, and division, becomes almost impossible. This failure is due to the different constraints imposed by the underlying hardware architecture and programming language. Since most of today's PC's CPU registers are 32-bit (4 bytes) and 64-bit (8 bytes) wide, they can only accommodate numbers of that length. [2] Additionally, data types in programming languages are kind of a culprit, for instance, in the foremost programming languages, an *int* data type can hold up to 32 bits, and a *long* data type can hold up to 64 bits.

Different algorithms and techniques were developed to solve the problem of arithmetic computation on big numbers; however, all these algorithms implement the same principles: they first convert big numbers from base-10 to base-2, then they execute bitwise operations on the bit level [3,4]. For instance, arithmetic addition can be performed using the bitwise logical operators OR and XOR. Such algorithms are of complexity $O(n)$, where $n$ is the total number of bits composing each of the big operands.

In this paper, we propose two new algorithms for handling arithmetic addition of big-integer numbers. The first one is a sequential algorithm designed to run on single-processor systems, and the second, is a parallel algorithm designed to run on multi-processor shared memory architecture systems. The aim behind these algorithms is to i) enhance the execution time and reduce the time complexity of current existing algorithms, and to ii) present a parallel implementation for multiple processors systems that drastically decreases that amount of time needed to perform addition on big-integer numbers. By implementing the suggested algorithm, certain type of applications such as cryptography, financial, astrological, mathematical, and scientific applications will be able to carry out computer arithmetic addition on numbers larger than 64 bits, at high speed.

Unlike other approaches that operate on single bits, the proposed algorithms operate on tokens or blocks of 60 bits each (18 digits) over a 64-bit CPU architecture. Both proposed algorithms share the same principle in that each two corresponding 60-bit tokens, one from every operand, are added the same way a child of four adds two numbers in the decimal base-10 numeral system using a pen and a paper. The sequential algorithm adds the given tokens sequentially from right to left, while the parallel algorithm assigns each two corresponding tokens to a particular processor to be added. Experiments showed outstanding results compared with those of the existing approaches.

## 2  Existing Solutions

Many programming libraries were developed to solve the problem of performing arithmetic calculations on big-integer numbers. Some of them are proprietary third party dynamic link libraries (DLL), either available for free or sold at a given cost, or shipped as a part of the programming language Application Programming Interface (API). For instance, the MS .NET Framework 4.0 provides the *BigInteger* class in the namespace *System.Numerics* [5]. The Java programming language provides another *BigInteger* class in the *java.math* package [6]. They both carry out arithmetic operations on big-integer numbers using bitwise operations [7]. They first convert the base-10 big-integer input to a base-2 binary representation, then they employ the bitwise operators OR and XOR to perform binary addition on string of bits.

The algorithm behind these libraries is of complexity $O(n)$, where *n* is the total number of bits constituting each operand. In terms of time efficiency, the number of times the basic operations OR and XOR is executed, is equal to the number of bits in the big-integer operands. Moreover, most of these libraries are not designed to work in a parallel fashion, but are to operate over single-processor systems.

Most of the publications and researches done to provide a solution for arithmetic addition on big-integer numbers tackle the problem from a hardware perspective and not from a software algorithmic perspective. For instance, Fagin [8] proposed an enhancement for the carry-look-ahead adder. His idea revolves around the use of a massively parallel computer with thousands of processors, each connected to a local memory and a communication network. By distributing the integers to be added among the processors, parallel prefix techniques are employed to rapidly add large numbers in a faster manner than if conventional machines were employed. Avizienis [9] proposed a representation schema for binary numbers used in fast parallel arithmetic. This schema revolves around replicating each input operands in order to eliminate the chain carry propagation during an arithmetic addition and subtraction.

## 3  The Proposed Sequential Algorithm

The sequential algorithm proposed in this paper is based on the same principle humans use to perform addition, in the decimal system, using a pencil and a paper. Generally speaking, inputs of big-integer numbers are cut into several smaller tokens, each made out of 60 bits (18 digits). Afterwards, each two corresponding tokens are treated as single units and aligned on top of each other; then they are added while handling appropriately the generated carries. In this approach, no conversion to base-2 is to occur; the computation is totally done in the base-10 decimal system

Below are the steps the sequential algorithm executes to add two big-integer numbers:

1. Two big-integer operands **a** and **b**, both of type *string* and possibly not of the same length, are fed to the algorithm. *AddBigInteger(a , b)*
2. Both *string* operands **a** and **b** are then parsed and divided, from right to left, into smaller chunks or tokens $t_i(p)$, where **i** is the token index and **p** is the operand to which $t_i$ belongs. Consequently, operand **a** = $t_{n-1}(a)$… $t_0(a)$ and operand **b** = $t_{m-1}(b)$… $t_0(b)$, where **n** and **m**, are the total number of tokens constituting each of the operands. The length of each single produced token $t_i$ is less than or equal to 18. (In the C# programming language, the largest integer data type is *long* (signed by default) which can store up to 19 digits or $2^{63}$=9223372036853775808. Since in mathematical addition there is always a potential arithmetic overflow, it is crucial to reserve 1 digit for a possible *carry*, resulting in 19-1=18 digits represented by 60 bits). The resulting tokens will be stored as *string*s in two arrays, each for a particular operand.
3. The tokens contained in the two arrays are to be converted from *string* to *long* data type. In other words, each single token, now representing an array element with a maximum length of 18 digits, is to be converted to an integer value of type *long*. The conversion is required because arithmetic addition cannot be performed on *string* types
4. Both arrays, now containing *long* type tokens, are aligned on top of each other. Starting from the rightmost token, each two corresponding tokens are added as in performing addition using a pencil and a paper: $t_i(c)$ =  $t_i(a)$ + $t_i(b)$ where $t_i(c)$ should be less than or equal to 18 digits; otherwise, the leftmost digit (the 19[th] digit) is truncated and added as a *carry* to the result of $t_{i+1}(a)$ + $t_{i+1}(b)$; $t_{i+1}$ is the next token on the left of the two tokens being currently added. It is worth noting that sometimes the length of $t_i(c)$ can be less than 18 digits. This is the case when $t_i(a)$ and $t_i(b)$ are the last leftmost tokens.

5. Finally, all the produced $t_i(c)$ are to be concatenated together to attain **result** = $t_{r-1}(c)$… $t_0(c)$. It is important to note that this algorithm can handle operands of different sizes, in a sense that excessive tokens, which should logically belong to the largest operand, are just appended to the final result. Figure 1 summaries the different steps performed by the sequential algorithm in order to add two operands **a** and **b**.

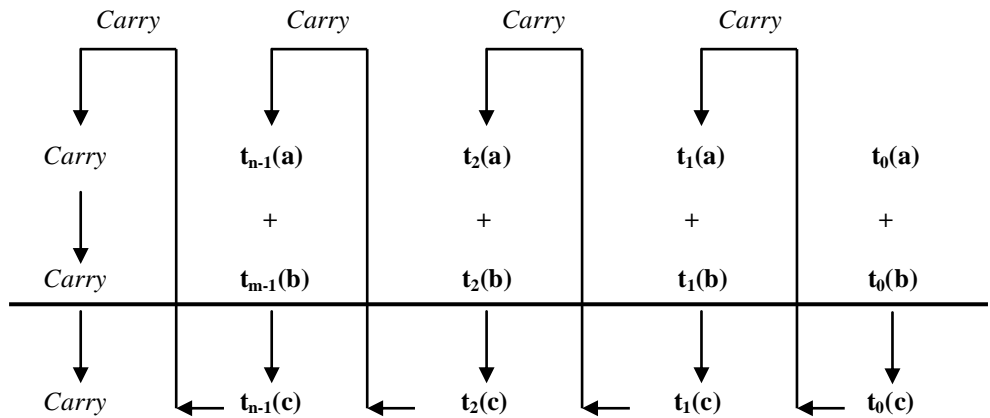

**Fig. 1.** Adding two big numbers using the proposed sequential algorithm

### 3.1 Implementation

Below is the code of the sequential algorithm implemented using MS C#.NET 2005 under the .NET Framework 2.0 and MS Visual Studio 2005.

```
private string AddBigInteger(string a, string b)
{
  long[] tokens_A = ParseOperand(a);
  long[] tokens_B = ParseOperand(b);

  int length = tokens_A.Length ;

  long[] result = new long[length];

  int i, j;
  for (i = length - 1, j = length - 1; j != -1; i--, j--)
  {
      result[i] = result[i] + tokens_A[i] + tokens_B[j];

      if (i != 0 && result[i].ToString().Length > 18)
      {
         result[i - 1] = 1;

         result[i] = result[i] % 1000000000000000000;
```

```
      }
    }

    return ConvertToString(result);
  }

  private long[] ParseOperand(string operand)
  {
     ArrayList list = new ArrayList();

     for (int i = 0; operand.Length > 18; i++)
     {
        list.Add(operand.Substring(operand.Length - 18));

        operand = operand.Substring(0, operand.Length - 18);
     }

     list.Add(operand);
     list.Reverse();

     long[] tokens = new long[list.Count];

     for (int j = 0; j < tokens.Length; j++)
     {
        tokens[j] = Convert.ToInt64(list[j]);
     }

     return tokens;
  }
```

### 3.2 Algorithm Complexity

The method AddBigInteger() contains one *for* loop whose body is executed *n* times, where *n* is the total number of tokens in the input operand. Each token is made out of 60 bits, which is equivalent to 18-digits in the decimal system. Ignoring the instructions outside the loop and taking into consideration the most costly instruction "*result[i] = result[i] + tokens_A[i] + tokens_B[j];*" as the basic operation, we get the following:

$$\sum_{i=0}^{n} 1 = n \text{ and thus this algorithm is of time complexity } O(n)$$

Since the basic operation is to executed *n* times regardless of the value of the input, we get $C_{Best}(n) = C_{Worst}(n) = C_{Average}(n) = n$

### 3.3 Comparing Algorithms in Theory

In this section, we will be comparing our proposed sequential algorithm with the already existing *java.math.BigInteger* algorithm from a time complexity perspective. Below is a code snippet extracted from the source code[7] of the *java.math.BigInteger*

```
// code segment that performs bit-by-bit addition
byte carry = 0;
for(int i = 0; i < n; i++)
{
    // data is the byte array holding the bits of the large number.

    byte sum = data1[i] + data2[i] + carry;
    carry  = sum >> 1;
    data_result[i] = (byte)(sum & 0xFF);
}
```

As illustrated in the above code, the body of the *for* loop is executed $n$ times, where $n$ is the total number of bits in the operands. Ignoring the instructions outside the loop and taking into consideration the most costly instruction "*byte sum = data1[i] + data2[i] + carry;*" as the basic operation, we get the following:

$$\sum_{i=0}^{n} 1 = n \text{ and thus this algorithm is of time complexity } O(n)$$

Since the basic operation is to executed *n* times regardless of the value of the input, we get $C_{Best}(n) = C_{Worst}(n) = C_{Average}(n) = n$

Even though the existing algorithms and our proposed one are both of complexity $O(n)$, the $n$ in our proposed algorithm is always 60 times smaller since each token is made out of 60 bits. For instance, in the java algorithm, a decimal big-integer number of length 13 digits (1 trillion = $10^{12}$) is represented in 40-bits. This requires the basic operation to be executed 40 times. If we double the length to 80-bits, the number of times the basic operation is executed is doubled to 80 times. On the other hand, in our proposed algorithm, a decimal big-integer number of length 13 digits (1 trillion = $10^{12}$), is represented in 1 token since the maximum size of the token is 18 digits. This requires the basic operation to be executed only 1 time. If we double the length to 26 digits, the number of times that basic operation is executed is doubled: 2 times. Table 1 shows the number of iterations executed by each algorithm in order to add two big-integer numbers of the specified length.

**Table 1.** Iterations executed by both algorithms

| Input Length (base-10) | Iterations Executed by the Existing Algorithms | Iterations Executed by our Proposed Sequential Algorithm |
|---|---|---|
| 1 digit | 1 bit → 1 iteration | 1 token → 1 iteration |
| 10 digits | 34 bits → 34 iterations | 1 token → 1 iteration |
| 32 digits | 64 bits → 64 iterations | 2 tokens → 2 iterations |
| 101 digits (Googol=$10^{100}$) | 333 bits → 333 iterations | 101/18=6 tokens → 6 iterations |

### 3.4 Comparing Algorithms in Practice

We will be comparing, in our tests, the execution time of the proposed sequential algorithm with the *System.Numercis.BigInteger* class included in MS .NET Framework 4.0, and the *java.math.BigInteger* class included in Java SE 1.6.

Below are two code segments that illustrate how to use the methods of the built-in classes *System.Numercis.BigInteger* and *java.math.BigInteger* in order to add two big-integer numbers using the C#.NET and the Java language.

```csharp
using System.Numerics;
public class BigIntegerTest_Csharp
{
    public static void Main(string args[])
    {
        String operandA = "12345678909876543211234567890987654321" ;
        String operandA = "12345678909876543211234567890987654321" ;
        BigInteger a = BigInteger.Parse(operandA) ;
        BigInteger b = BigInteger.Parse(operandB) ;

        BigInteger results = BigInteger.Add(a, b);
        Console.WriteLine(results.ToString());
    }
}
```

```java
import java.math.BigInteger;
public class BigIntegerTest_Java
{
    public static void main(String args[])
    {
```

```
        String operandA = "12345678909876543211234567890987654321" ;
        String operandA = "12345678909876543211234567890987654321" ;
        BigInteger a = new BigInteger(operandA) ;
        BigInteger b = new BigInteger(operandB) ;

        System.out.print("" + a.add(b)) ;
    }
}
```

As a testing platform, we are using a desktop IBM-compatible PC with Intel Core single core processor with 1.66 MHz clock speed, 256KB of cache, and 512MB of RAM. The operating system used is MS Windows XP Professional SP2.

It is worth noting that the execution time obtained for all different algorithms is an average time obtained after five consecutive runs of the same test.

**Table 2.** Test cases

| Test Case | Operands | Value | Value Length |
|---|---|---|---|
| 1 | A | X | 20,000 base-10 digits |
| 1 | B | Y | 20,000 base-10 digits |
| 2 | A | X | 100,000 base-10 digits |
| 2 | B | Y | 100,000 base-10 digits |
| 3 | A | X | 500,000 base-10 digits |
| 3 | B | Y | 500,000 base-10 digits |
| 4 | A | X | 1000,000 base-10 digits |
| 4 | B | Y | 1000,000 base-10 digits |

**Table 3.** Results obtained from the .NET class

| Test Case | Operation | Results | Execution Time in Seconds |
|---|---|---|---|
| 1 | A+B | X+Y | 1.14 |
| 2 | A+B | X+Y | 18.02 |
| 3 | A+B | X+Y | 529.89 |
| 4 | A+B | X+Y | 2541.55 |

**Table 4.** Results obtained from the Java class

| Test Case | Operation | Results | Execution Time in Seconds |
|---|---|---|---|
| 1 | A+B | X+Y | 0.46 |
| 2 | A+B | X+Y | 13.16 |
| 3 | A+B | X+Y | 320.22 |
| 4 | A+B | X+Y | 1495.78 |

**Table 5.** Results obtained from our sequential algorithm

| Test Case | Operation | Results | Execution Time in Seconds |
|---|---|---|---|
| 1 | A+B | X+Y | 0.10 |
| 2 | A+B | X+Y | 2.23 |
| 3 | A+B | X+Y | 69.45 |
| 4 | A+B | X+Y | 327.90 |

From the obtained results delineated in tables 2-5, it is obvious that our sequential algorithm outsmarted all other algorithms in all different test cases. When big-integer numbers were respectively 20,000 and 100,000 in length, our algorithm beat the .NET and Java classes by few mere seconds. However, when numbers became as large as 500,000 digits, our algorithm surpassed the Java class by around 250 seconds (4 minutes), and the .NET class by around 460 seconds (7.6 minutes). Additionally, our proposed algorithm showed impressive results as compared to its rivals when the length of operands reached the 1,000,000 digits: it surpassed the Java class by around 1168 seconds (19.4 minutes), and the .NET class by around 2214 seconds (36.9 minutes).

### 3.5 Experiments' Analysis & Conclusion

The sequential algorithm showed a significant improvement over other existing approaches. It outperformed the .NET and Java built-in classes by several seconds. This gap exponentially increased as the length of the big-integer operands became larger. This speed improvement is due to the reduction of the input size *n* in $O(n)$. The .NET, Java, and our proposed algorithm are all of complexity $O(n)$. However, the *n* in the .NET and Java algorithms represents the total number of bits in each operand; while in our proposed algorithm, it represents the total number of tokens in each operand. For instance, the decimal number 999999999999999999 (18 digits) is represented in base-2 as 110111100000101101101011001110100111011000111111111111111111 (60 bits). This makes *n=60* and thus the basic operation is executed 60 times. On the other hand, in our proposed algorithm the whole decimal number 999999999999999999 is treated as a single unit token, making *n=1*, and thus the basic operation is executed only 1 time. When no carries are generated during the execution of our algorithm, then the algorithm is in its best-case, and its time efficiency is 60 times faster than the other algorithms. However, when several carries are produced, then various extra operations are to be executed; a fact that imposes further processing overhead, and increases the computation time of our algorithm. On the average, the time efficiency of our algorithm is 6 to 8 times faster than any other algorithm as demonstrated in tables 2-5.

## 4  The Proposed Parallel Algorithm

The parallel algorithm proposed in this paper is a multithreaded parallel algorithm designed to be executed over multi-processor shared memory architecture. It is based on the principle of performing arithmetic addition as humans perform addition in the decimal system using a pencil and a paper. Ordinarily, the algorithm starts by breaking down big-integer numbers into blocks or tokens of 60 bits each. Then addition starts in a sequence of multiple iterations. On the first iteration, each two corresponding tokens are assigned to a particular thread, which will then add them using a particular microprocessor, while the generated carries from each thread are stored in a shared array. On the second iteration, previous carries stored in the shared array are added properly to the previous result. Iterations continue until no more carries are generated from a previous iteration.

Below are the steps the parallel algorithm execute to add two big-integer numbers:

1.  Two very large numbers operand **a** and operand **b**, both of *string* type and possibly not of the same length, are fed to the algorithm. *AddBigInteger_Parallel(a , b)*
2.  Both *string* operands **a** and **b** are then parsed and divided from right to left into smaller chunks or tokens $t_i(p)$, where **i** is the token index and **p** is the operand to which $t_i$ belongs. Consequently, operand $a = t_{n-1}(a)… t_0(a)$ and operand $b = t_{m-1}(b)… t_0(b)$, where **n** and **m** are the total number of tokens constituting each of the operands. The length of each single produced token $t_i$ is less than or equal to 18 (In the C# programming language, the largest integer data type is *long* (signed by default), and which can store up to 19 digits or $2^{63}$=9223372036853775808. Since in mathematical addition there is always a potential arithmetic overflow, it is crucial to reserve 1 digit for a possible *carry*, resulting in 19-1=18 digits represented by 60 bits). The resulting tokens will be stored as *string* in two arrays, each for a particular operand.
3.  The tokens contained in the two arrays are to be converted from *string* to *long* data type. In other words each single token, now representing an array element with a maximum length of 18 digits, is to be converted to an integer value of type *long*. This conversion is required because arithmetic addition cannot be performed on *string* types
4.  Each processor $p_i$ in a multiprocessor system is assigned two tokens, one from each operand. Therefore, the processor $p_i$ is assigned tokens $t_i(a)$ and $t_i(b)$ with the purpose of calculating $t_i(c) = t_i(a) + t_i(b)$. For instance, $p_0$ will calculate $t_0(c)$, $p_1$ will calculate $t_1(c)$, $p_2$ will calculate $t_2(c)$ and so on and so forth. We are to assume that the number of

    processor is equal to the number of tokens; otherwise, tokens are distributed equally among processors. For instance, if the number of processors is half the number of tokens, each processor will be assigned 4 tokens (2 from each operand) to be calculated as in sequential approach. $t_i(c) = t_i(a) + t_i(b)$ and then $t_{i+1}(c) = t_{i+1}(a) + t_{i+1}(b)$

5. The C*arry* generated from each $t_i(c)$ is handled using multiple processing iterations, and a shared array called **carries[0...n-1]** is used to store all the produced carries. For that reason, we have added a new variable called **T** as in $t_i(c,T)$ to represent the iteration into which $t_i(c)$ is being calculated. **T=1** is the first iteration and **T=n** is the nth iteration. In this approach, if a carry surfaced after calculating $t_i(c,1)$, **carries[i+1]** is set to 1. It is **i+1** so that on the next iteration **T=2**, **carries[i+1]** will be correctly added to the previously calculated $t_{i+1}(c,1)$. Likewise, if another carry surfaced from $t_i(c,2)$, **carries[i+1]** is set to 1 overwriting any previous value**.** Consequently, on the next iteration **(T=3) carries[i+1]** will be correctly added to $t_{i+1}(c,2)$. This will keep on looping until no more carries are generated (array **carries[0...n-1]** contains no 1's). As an example, if on the first iteration **(T=1)**, a carry is generated from $t_4(c,1)$**,** then **carries[5]** is set to 1, $p_5$ (processor 5) starts a second iteration **(T=2)** in an attempt to calculate $t_5(c,2) = t_5(c,1) + $ **carries[5]**. In the meantime, all other $p_i$, where **carries[i]=0**, will refrain from executing. If after **T=2** no carries was generated, the loop process stops.

6. Finally, all the $t_i(c)$ produced after many iterations are to be concatenated together: **result** = $t_{n-1}(c)$… $t_0(c)$. Figure 2 summaries the different steps performed by the parallel algorithm in order to add two operands **a** and **b**.

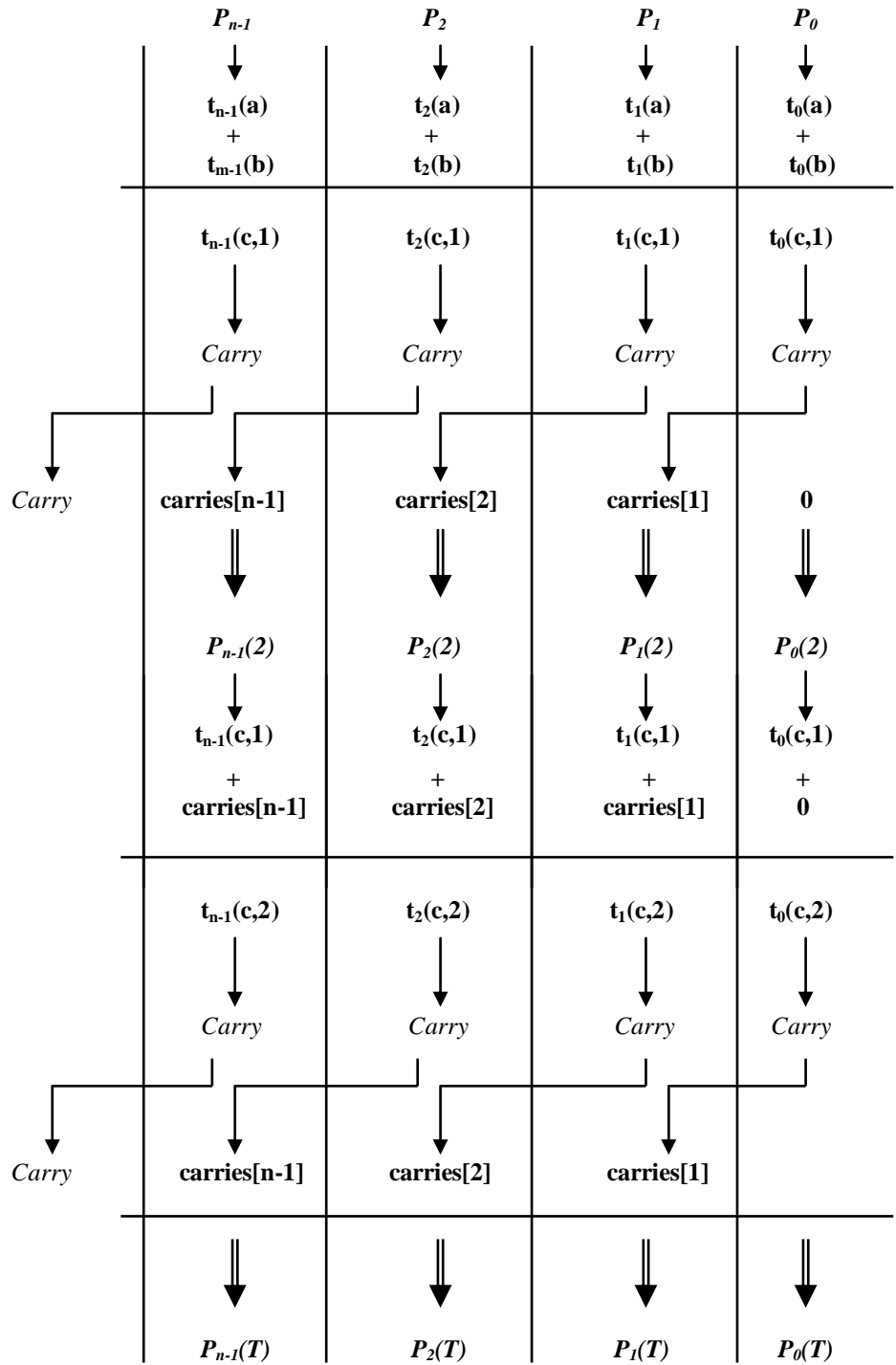

**Fig. 2.** Adding two big numbers using the proposed parallel algorithm

**4.1 Implementation**

Below is the code of the proposed parallel algorithm implemented in MS C#.NET 2005 under the .NET Framework 2.0 and MS Visual Studio 2005. It uses classes and methods from *System.Threading* namespace to create, destroy and execute threads. All threads can read and write to a shared memory space where tokens, carries, flags and other variables are stored and shared.

```csharp
long[] tokens_A ;
long[] tokens_B ;

long[] result;
int[] carries;

int numberOfProcessors;

int sharedIndex;
int terminatedThreads=0;
int T=1;

Thread[] threads;

public void AddBigInteger_Parallel(string a, string b)
{
   tokens_A = ParseOperand(a, 18);
   tokens_B = ParseOperand(b, 18);

   result = new long[tokens_A.Length];

   carries = new int[tokens_A.Length]; // By default the array carries is populated with 0s

   numberOfProcessors = GetNumOfProcessors();

   threads = new Thread[numberOfProcessors];

   CreateThreads();
}

private void CreateThreads()
{
   sharedIndex = numberOfProcessors;

   for (int i = 0; i < numberOfProcessors; i++)
   {
      threads[i] = new Thread(new ThreadStart(Process));
      threads[i].Start();
   }
}

private void Process()
{
   int index = sharedIndex--; // index is private to every thread
```

```
     if(T==1) // First iteration
     {
          result[index] = tokens_A[index] + tokens_B[index];
     }
     else result[index] = carries[index] + result[index];

     if (index != 0) // not the leftmost token
     {
        if (result[index].ToString().Length > 18) // a carry was generated
        {
           carries[index - 1] = 1;

           result[index] = result[index] % 1000000000000000000; // discarding the carry
        }
        else carries[index - 1] = 0;
     }

     terminatedThreads++;

     IsProcessingDone();
  }

  private void IsProcessingDone()
  {
     if (terminatedThreads == numberOfProcessors)
     {
        if (AreMoreCarries())
        {
           T++ ;
           CreateThreads(); // Creates new set of threads in the next iteration
        }
        else DisplayResults();
     }
  }

  private bool AreMoreCarries()
  {
     for (int i = 0; i < carries.Length; i++)
     {
        if (carries[i] == 1)
           return true;
     }

     return false;
  }

  private string DisplayResults()
  {
     return ConvertToString(result);
  }

  private long[] ParseOperand(string operand)
  {
     ArrayList list = new ArrayList();
```

```
    for (int i = 0; operand.Length > 18; i++)
    {
       list.Add(operand.Substring(operand.Length - 18));

       operand = operand.Substring(0, operand.Length - 18);
    }
    list.Add(operand);
    list.Reverse();

    long[] tokens = new long[list.Count];

    for (int j = 0; j < tokens.Length; j++)
    {
       tokens[j] = Convert.ToInt64(list[j]);
    }

    return tokens;
}
```

### 4.2 Algorithm Complexity

The method Process() is called by each thread running on a particular processor. The method has no *for* loop at all, it is basically 1 instruction executed on each processor to add two corresponding tokens together. Therefore, and taking into consideration the most costly instruction "*result[index] = tokens_A[index] + tokens_B[index];*" as the basic operation, we get the following:

**Case 1:** if tokens are equally distributed among processors, *n* tokens are assigned to *n* processors. The basic operation is executed only 1 time by each processor and hence the time complexity of the algorithm is *O(1)*

**Case 2:** if tokens are not equally distributed among processors, *n* tokens are assigned to *m* processors, where *m < n*. As a result, the basic operation might be executed multiple times by the same thread and processor. For instance, if n=100(100 tokens) and m=50(50 processors), the basic operation is executed 2 times by each processor, and thus the algorithm is of complexity *O(2)*

**Conclusion:** From both cases we can conclude that the algorithm is of complexity *O(n/m)*, where *n* is the total number of tokens and *m* is the total number of processors.

The best-case efficiency is when no carries are generated after the first iteration; hence, achieving the best performance where $C_{Best}(n)=1$, that is, each processor executes the basic operation only one time. The worst-case efficiency is when a new carry is generated after each iteration, this would require *n-1* iterations in order to propagate and add all the carries. Thus $C_{Worst}(n)=n-1$. Consequently, the average-case efficiency is $C_{Average}(n)=(n-1)/2$

### 4.3 Comparing Algorithms in Theory

In this section, we will be comparing our proposed sequential and parallel algorithm from a time complexity perspective. The sequential algorithm is of best-case complexity $O(n)$; while the parallel algorithm is of best-case complexity $O(1)$. This means that as long as the number of tokens $n$ is equal to the number of processors $m$, the parallel algorithm is faster by $n$ times. For instance, a decimal big-integer number of length 1000 digits (1 thousand) is represented by 55 tokens (1000/18=55), where 18 is the length of each token. In the sequential algorithm this would require the basic operation to be executed 55 times; whereas, in the parallel algorithm, this would require the basic operation to be executed only 1 time, but under one condition that the algorithm is executed over 55 processors.

Table 6 shows the number of times the basic operation is executed by our two proposed algorithms, taking into consideration that the number of tokens is equal to the number of processors → $n = m$

Table 7 shows the number of times the basic operation is executed by our two proposed algorithms, taking into consideration that the number of processors is half the number of tokens → $m = n/2$

**Table 6.** Number of times the basic operation is executed when $n=m$

| Input Length in base-10 | Number of Times the Basic Operation is Executed in the Sequential Algorithm | Number of Times the Basic Operation is Executed in the Parallel Algorithm |
|---|---|---|
| 1000 digit (55 tokens) | 55 times | 1 time |
| 100,000 digits (5555 tokens) | 5555 times | 1 time |
| 1,000,000 (55555 tokens) | 55555 times | 1 time |

**Table 7.** Number of times the basic operation is executed when $m = n/2$

| Input Length in base-10 | Number of Times the Basic Operation is Executed in the Sequential Algorithm | Number of Times the Basic Operation is Executed in the Parallel Algorithm |
|---|---|---|
| 1000 digit (55 tokens) | 55 times | 2 times |
| 100,000 digits (5555 tokens) | 5555 times | 2 times |
| 1,000,000 (55555 tokens) | 55555 times | 2 times |

### 4.4 Comparing Algorithms in Practice

In this section, a comparison of the execution time between the sequential and the parallel algorithm is undertaken using a desktop IBM-compatible PC with 4 processors of type Intel Core single core with 1.8 MHz clock speed, 512KB of cache, and 2GB of RAM. The operating system used is MS Windows Server 2003 SP1.

It is important to note here that the execution time obtained for all different algorithms is an average time obtained after five consecutive runs of the same test.

**Table 8.** Test cases

| Test Case | Operands | Value | Value Length |
|---|---|---|---|
| 1 | A | X | 20,000 base-10 digits |
| 1 | B | Y | 20,000 base-10 digits |
| 2 | A | X | 100,000 base-10 digits |
| 2 | B | Y | 100,000 base-10 digits |
| 3 | A | X | 500,000 base-10 digits |
| 3 | B | Y | 500,000 base-10 digits |
| 4 | A | X | 1000,000 base-10 digits |
| 4 | B | Y | 1000,000 base-10 digits |

**Table 9.** Results obtained from our sequential algorithm

| Test Case | Operation | Results | Execution Time in Seconds |
|---|---|---|---|
| 1 | A+B | X+Y | 0.09 |
| 2 | A+B | X+Y | 2.01 |
| 3 | A+B | X+Y | 62.12 |
| 4 | A+B | X+Y | 310.34 |

**Table 10.** Results obtained from our parallel algorithm

| Test Case | Operation | Results | Execution Time in Seconds |
|---|---|---|---|
| 1 | A+B | X+Y | 0.04 |
| 2 | A+B | X+Y | 0.80 |
| 3 | A+B | X+Y | 20.1 |
| 4 | A+B | X+Y | 93.12 |

## 4.5 Experiments' Analysis & Conclusion

The results delineated in tables 8-10 show that the parallel algorithm outperformed the sequential algorithm by an average factor of 3.2. At the beginning, when operands were respectively 20,000 and 100,000 in length, the difference was not that evident. However, when numbers became larger, the gap increased and the execution time was speeded up by around 320%. Since 4 processors were only used ($m=4$), each processor was assigned $n/4$ tokens from each operand. In best-case, no carries are to be generated and thus the basic operation is executed

*n/4* times. In worst-case, *(n/4)-1* carries are to be generated and thus the basic operation is executed *n/4 + (n/4)-1*

# 5  Future Work

Future research can improve upon our proposed algorithms so much so that other arithmetic operations such as subtraction, multiplication, and division are added. Besides, a distributed version of the same algorithms could be designed so that it can be executed over a network of regular machines, making the implementation less expensive and more scalable.

## Acknowledgments

We thank Mrs. Henriette Skaff in the Department of Languages and Translation at AUST for her help in editing this article.